\documentclass[letter,twocolumn]{jpsj2} 

\usepackage{color}

\def\runauthor{S. {\sc Morohoshi} and Y. {\sc Fukumoto}}
\title{Superconductivity in the Three-Fold Charge-Ordered Metal of the Triangular-Lattice Extended Hubbard Model}

\author{Shotaro \textsc{Morohoshi}
and Yoshiyuki \textsc{Fukumoto}
}

\inst{Department of Physics, Faculty of Science and Technology, 
        Tokyo University of Science, Noda, Chiba 278-8510 }
\recdate{\hspace{3cm}}

\abst{
The quarter-filling extended Hubbard model on the triangular lattice is studied
to explore pairing instability in the three-fold charge-ordered (CO) metal.
We derive a second-order strong-coupling effective Hamiltonian of doped carriers into the three-fold CO insulator at electron density of $n=2/3$,
and then study the $f$- and $d_{xy}$-wave superconductivities down to $n=1/2$ by using the BCS mean-field approximation.
It is found that the triplet $f$-wave pairing is more stable than the $d_{xy}$-wave one.
We also discuss that this coexisting state of the charge ordering and superconductivity is possible to have critical temperature $T_{\rm c}\sim 0.01 t$.
}

\kword{triangular lattice, three-fold charge order, extended Hubbard model, superconductivity, BCS mean-field approximation, organic conductor, $\theta$-(BEDT-TTF)$_2$X}

\begin{document}
\maketitle


Motivated by experimental investigations on $\theta$-type organic conductors~\cite{mori} 
and Na$_x$CoO$_2$$\cdot y$H$_2$O,~\cite{takada}
the possible relevance of charge ordering and superconductivity has been studied extensively
on the basis of the triangular-lattice extended Hubbard model with an additional nearest-neighbor Coulomb repulsion $V$.~\cite{watanabe,watanabe2,tanaka,onari,kuroki}.

Charge-fluctuation mediated superconductivity in the vicinity to a CDW phase is one of main topics in those investigations.
Tanaka {\it et al.} applied the RPA method to the triangular-lattice extended Hubbard model,
having the superconductivity in Na$_x$CoO$_2$$\cdot y$H$_2$O in their mind.~\cite{tanaka}
They found that $V$ enhances the charge fluctuation at ${\mib q}=(2\pi/3,2\pi/3)$ and triplet next-nearest-neighbor $f$-wave superconductivity is induced.
Onari {\it et al.} obtained phase diagram by using the FLEX approximation.~\cite{onari}
They found that $f$-wave pairing adjacent to a CDW phase is stable for intermediate filling and large $V$. 
These results have been paid attentions, because triplet superconductivity was considered to be stabilized only under some special conditions.

\begin{figure}[b]
\begin{center}
\includegraphics[width=.80\linewidth]{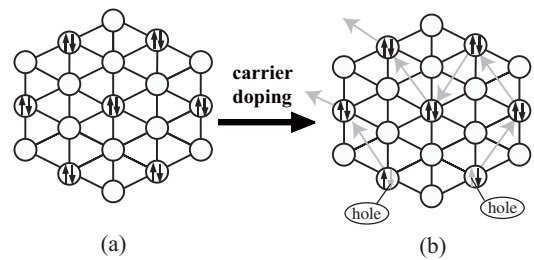}
\end{center}
\caption{
Schematic representations of (a) the three-fold CO insulator realizing for $V>U/3$ at $n=2/3$ and (b) the three-fold CO metal 
which is obtained by hole doping into the CO insulator.
Second-order perturbation processes make doped holes be itinerant within A-sublattice, which is schematically represented by the gray arrows.
}
\label{fig:f1}
\end{figure}

Due to frustration effects inherent to the triangular-lattice, a variety of charge-ordering patterns have been considered in this system.~\cite{hotta,watanabe,watanabe2,udagawa,kaneko,kuroki}
Watanabe {\it et al.} studied the triangular-lattice extended Hubbard model at $n=2/3$ by the variational Monte-Carlo (VMC) method, and found two kinds of CO states, 
three-fold bipolaronic CO state and antiferromagnetic CO state, in the $U$-$V$ phase diagram.~\cite{watanabe}.
In the three-fold CO state, A-sublattice sites are doubly occupied and B- and C-sublattices have no electrons, as shown in Fig.~\ref{fig:f1}(a).
In the antiferromagnetic CO state, there exist no electrons on A-sublattice.
Each of B- and C- sublattice sites is singly occupied, and the spins are in honeycomb-type antiferromagnetic arrangement.
Ground-state energies per site of the three-fold and antiferromagnetic CO states 
are, respectively, $U/3$ and $V$ in the strong-coupling limit, and thus we expect the transition point $V\simeq U/3$.
And then they studied an anisotropic-triangular-lattice extended Hubbard model at quarter filling, 
which is a model for $\theta$-(BEDT-TTF)$_2$X, by the VMC method.~\cite{watanabe2}.
They found that there appears three-fold CO metallic phase rather than insulating stripe CO phases if the anisotropy in the intersite repulsions is not so large.

Organic conductors $\theta$-(BEDT-TTF)$_2$X have been known to show superconductivity when X$=$I$_3$.
Watanabe {\it et al.} speculated that $\theta$-(BEDT-TTF)$_2$I$_3$ is located in the vicinity of the para metal phase in their ground-state phase diagram,\cite{watanabe2} and thus it is a weak coupling system.
Aside from the description of organic conductors by a realistic theoretical model, there arises a fundamental question: 
Is the extended Hubbard model possible to show superconductivity in intermediate and strong coupling regions?
When Coulomb interactions are increased, then a charge ordering is expected to be more favored.
If the CO state has a charge excitation gap, then superconductivity is never obtained. 
On the other hand, if the CO state is metallic, then charge carriers are possible to be condensed into a superconducting state.
Such superconducting state is inevitably coexisting with the charge ordering.

The purpose of this paper is to explore theoretically the above-mentioned coexisting state in the triangular-lattice extended Hubbard model with quarter-filled band.
To be precise, we study pairing instability in the three-fold CO metallic phase.
Unfortunately, it has not been reported that the three-fold CO metallic phase is realized in organic conductors.
We hope that the three-fold CO metallic phase is found experimentally in the near future, which opens new windows to the physics of charge ordering and superconductivity.

In order to study the present issue, we use the extended Hubbard model on the isotropic triangular-lattice,
\begin{equation}
   H= -t \sum_{\langle i,j \rangle \sigma}(c^{\dagger}_{i\sigma}c_{j\sigma}+\text{h.c.})
   +U \sum_i n_{i\uparrow}n_{i\downarrow}+ V\sum_{\langle i,j \rangle}n_{i}n_{j},
\label{eq.1}
\end{equation}
where $\langle i,j \rangle$ denote nearest-neighbor pairs of lattice points.
Hereafter, we assume the strong-coupling limit, $t\ll U,\;V$, in our calculations. For the stability of the three-fold charge ordering,  we also assume $U<3V$.

Our starting point is the three-fold CO insulator at $n=2/3$, which is schematically shown in Fig.~\ref{fig:f1}(a).
If we make hole doping to this insulator, the three-fold CO metal is obtained.
Note that the idea of carrier doping to an insulator is similar to the high-$T_{\rm c}$ cuprates.~\cite{fukuyama}
Figure \ref{fig:f1}(b) illustrates the strong-coupling picture of how doped holes propagate in the CO background: 
a hole can hop from an A-sublattice site to another A-sublattice site next to the first one via the second-order perturbation process of the bare hopping $t$.

Some readers may wonder if the hole doping into a bipolaronic CO insulator collapses the charge order drastically, like doped antiferromagnets.~\cite{Anderson}
This issue has been studied by several authors in the linear-chain,\cite{ohta} two-leg ladder,\cite{vojta} and square lattice.\cite{ohta,fukumoto}
Those studies indicated that charge order is robust against hole doping in the intermediate- and strong-coupling regions.
As for the triangular lattice treated in this paper,
Hotta {\it et al.} recently studied the $t$-$V$ model for spinless fermions at the density $\rho \sim 1/3$.~\cite{hotta} 
They confirmed that the three-fold charge ordering survives even at $\rho >1/3$ and the charge excitation gap closes there.
Note that charge order can be stabilized only by the Coulomb repulsion, although both the hopping and Coulomb repulsion are required to stabilize antiferromagnetic order.
The difference between CO and antiferromagnetic states may be coming from this fact.

Now we turn to our theoretical analysis.
In order to identify an ordering stabilized in the doped three-fold CO system, we calculate the second-order effective Hamiltonian within the two-hole approximation.~\cite{onozawa}
The obtained effective Hamiltonian is an triangular-lattice extended Hubbard model with additional nearest-neighbor interaction and correlated hopping terms:
\begin{align}
 H_{\rm{eff}}
   &=T_1\sum_{\langle\ell,\ell^{\prime}\rangle}h^a_{\ell,\ell^{\prime}}
   +V_0\sum_{\ell}n_{\ell\uparrow}^a n_{\ell\downarrow}^a
   \nonumber \\
   &\hspace{3mm}+V_1\sum_{\langle\ell,\ell^{\prime}\rangle} n_{\ell}^a n_{\ell^{\prime}}^a
   +\tilde{T}_1\sum_{\langle\ell,\ell^{\prime}\rangle}\sum_{\ell^{\prime\prime}\in {\cal{T}}_{\ell,\ell^{\prime}}}
   n^a_{\ell^{\prime\prime}}h^a_{\ell,\ell^{\prime}},
   \label{eq.eff}
\end{align}
where $\ell$ denotes an A-sublattice site, $\langle \ell,\ell^{\prime}\rangle$ a nearest-neighbor pair of A-sublattice sites, 
${\cal{T}}_{\ell,\ell^{\prime}}$ a set of A-sublattice sites which form regular triangles together with the lattice points $\ell$ and $\ell^{\prime}$,
$h^a_{\ell,\ell'}=\sum_{\sigma}(a^{\dagger}_{\ell\sigma}a_{\ell'\sigma}+\mbox{h.c.})$ with a hole annihilation operator $a_{\ell\sigma}=c^{\dagger}_{\ell\sigma}$ (hole-particle transformation),
$n_{\ell\sigma}^a=a^{\dagger}_{\ell\sigma}a_{\ell\sigma}, n_{\ell}^a=\sum_{\sigma}n_{\ell\sigma}^a$. The matrix elements in $H_{\rm{eff}}$ are given by
\begin{equation}
   T_1=\frac{2t^2}{V}\frac{1}{4-w},
\label{eq:t1}
\end{equation}
\begin{equation}
   V_0=U-\frac{12t^2}{V}\left(-\frac{1}{4}+\frac{2}{3-w}-\frac{4}{4-w}+\frac{4}{5-w}\right), 
\label{eq:v0}
\end{equation}
\begin{equation}
   V_1=-\frac{4t^2}{V}\left(\frac{1}{12}+\frac{1}{3-w}-\frac{4}{4-w}+\frac{3}{5-w}\right), 
\label{eq:v1}
\end{equation}
\begin{equation}
   \tilde{T}_1=\frac{t^2}{V}\left(\frac{1}{3-w}-\frac{1}{4-w}\right),
\label{eq:t2}
\end{equation}
where $w=U/V$.

\begin{figure}[b]
\begin{center}
\includegraphics[scale=0.42]{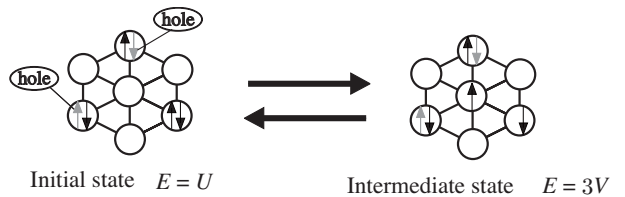}
\end{center}
\caption{Schematic representation of a second-order perturbation process by a bare hopping term,
where the black and gray arrows, respectively, represent electrons and holes.
This process is the origin of the nearest-neighbor attraction between two holes. 
The energy of the initial state is $U$ because of the double occupancy,
and that of the intermediate state is $3V$ because of the intersite repulsion $V$.
(See text in detail.)
}
\label{f2}
\end{figure}

Let us examine each of the matrix elements.
First, $T_1$ makes a hole hop from an A-sublattice site to another nearest-neighbor A-sublattice site, as described before.
Secondly, eq.~(\ref{eq:v0}) tells us that the on-site interaction $V_0$ is repulsive and $V_0\simeq U$.
Thirdly, we find the nearest-neighbor interaction $V_1$ being attractive in eq.~(\ref{eq:v1}), and thus it could be a driving force of superconductivity.
The origin of this attraction is as follows. 
We suppose that there exists a nearest-neighbor pair of doped holes, and consider a process in which
an electron hops from an A-sublattice site to a B-sublattice (or C-sublattice) site next to the doped hole pair (see Fig.~\ref{f2}). 
The energy difference between the initial and intermediate states is $3V-U$, which vanishes in the limit $U/V\rightarrow 3$.
Such small energy difference is never obtained for more distant doped hole pair, so the nearest-neighbor interaction becomes attractive.
Finally, the matrix element of the correlated hopping, $\tilde{T}_1$, is positive. 
It has been known that correlated hopping induces superconductivity depending on positive and negative of its matrix element.~\cite{onozawa,hirsch}
As shown later, the positive $\tilde{T}_1$ actually stabilizes the $f$- and $d_{xy}$-wave pairings.

We here derive the dispersion relation of free holes by neglecting the interaction terms in $H_{\rm{eff}}$.
Making the Fourier transformation 
$a_{\mib{k}\sigma}=N^{-1/2} \sum_{\ell}e^{i{\mib k}\cdot{\mib r}_\ell}a_{\ell\sigma}$, 
where $N$ is the total number of A-sublattice sites, we obtain the dispersion relation
$\epsilon_k=2T_1[\cos k_y+2\cos(\sqrt{3}k_x/2)\cos(k_y/2)]$,
where the unit of length is the lattice constant of A-sublattice.
The band width is given by $9T_1$.

\begin{figure}[h]
\begin{center}
\includegraphics[scale=0.35]{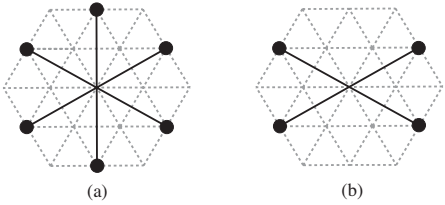}
\end{center}
\caption{Schematic representations of next-nearest-neighbor (a) $f$-wave and (b) $d_{xy}$-wave symmetries in real space. 
In ${\mib k}$-space, the gap function changes its sign six (four) times for $f$-wave ($d_{xy}$-wave). The dotted lines represent the original lattice. }
\label{f3}
\end{figure}

We turn to the study of superconductivity on the basis of the obtained effective Hamiltonian. 
Here we consider two types of pairing, $f$-wave and $d_{xy}$-wave shown in Fig~\ref{f3}. 
We define the off-diagonal mean-field parameter as
$\Lambda({\mib \rho})\equiv \langle a_{\ell\uparrow}a_{\ell+{\mib \rho}\downarrow}\rangle$, where
$\mib{\rho}$ represent nearest-neighbor vectors of A-sublattice.
The $f$- and $d_{xy}$-wave order parameters satisfy the symmetry relations
$\Lambda({\mib \rho})=-\Lambda(-{\mib \rho})$ and $\Lambda({\mib \rho})=\Lambda(-{\mib \rho})$, respectively.
We also denote the absolute value of $\Lambda({\mib \rho})$ as $\Lambda$.

Applying the BCS mean-field approximation to $H_{\rm eff}$ and using the Bogoliubov transformation, 
we obtain the diagonalized mean-field Hamiltonian 
$H_{\rm mf}=\sum_{k,\sigma}E_k\alpha^{\dagger}_{k\sigma}\alpha_{k\sigma}+\sum_k(\xi_k-E_k)+\gamma\Lambda^2$,
where $\alpha_{k\sigma}$ denotes the annihilation operator of a quasi particle and $E_k=\sqrt{\xi_k^2+\Delta^2f_k^2}$ with $\xi_k=\epsilon_k-\mu$,
\begin{equation}
   f_k=
   \begin{cases}
   \dfrac{2}{3}\sin\dfrac{k_y}{2}\left(\cos \dfrac{\sqrt{3}k_x}{2}
 -\cos\dfrac{k_y}{2}\right) & \mbox{for $f$-wave} \\
   \sin \dfrac{\sqrt{3}k_x}{2}\sin\dfrac{k_y}{2} & \mbox{for $d_{xy}$-wave}
\end{cases},
\end{equation}
and $\Delta=\gamma\Lambda$. The pairing interaction $\gamma$ is defined by
\begin{equation}
\label{eq:gam}
   \gamma=
   \begin{cases}
        6(4\tilde{T}_1-V_1)
      & \mbox{for $f$-wave} \\
        4(2\tilde{T}_1-V_1)
      & \mbox{for $d_{xy}$-wave} 
\end{cases}.
\end{equation}
Minimizing the free energy with respect to the order parameter $\Delta$,
we obtain the following mean-field equation,
\begin{equation}
    1= \frac{\sqrt{3}\gamma}{2\pi^2}\int_{0}^{\pi}\int_{0}^{\frac{2\pi}{\sqrt{3}}}dk_xdk_y
    \frac{f_k^2}{2E_k}\tanh \left(\frac{E_k}{2k_{\rm B}T}\right).
    \label{eq:mfeq1}
\end{equation}
The chemical potential $\mu$ is determined by the condition $n_{\rm h}=N^{-1}\sum_{\ell}\langle n_{\ell}^a\rangle$, which turns out to be
\begin{equation}
    n_{\rm h}=\frac{\sqrt{3}}{2\pi^2}\int_0^{\pi}\int_{0}^{\frac{2\pi}{\sqrt{3}}}\left[1-\frac{\xi_{k}}{E_k}\tanh\left(\frac{E_k}{2k_{\rm B}T}\right)\right]dk_{x}dk_{y}.
    \label{eq:mfeq2}
\end{equation}
In eqs. (\ref{eq:mfeq1}) and (\ref{eq:mfeq2}),
we have transformed the integration range in the $\mib{k}$-space from
the hexagonal first Brillouin zone of the triangular-lattice to the rectangular region.

\begin{figure}[t]
\begin{center}
\includegraphics[width=.65\linewidth]{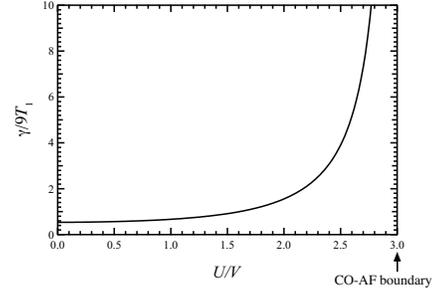}
\end{center}
\caption{Dependence of the pairing interaction $\gamma$ on $U/V$ for the $f$-wave pairing.
The CO-AF boundary is $(U/V)_{\rm c}=3$. 
}
\label{f4}
\end{figure}

In Fig.~\ref{f4}, we show the $U/V$ dependence of the pairing interaction $\gamma$, divided by the band width $9T_1$, for the $f$-wave pairing.
We obtain larger $\gamma$ when we get closer to the CO-AF boundary, $U/V=3$.
The value of $\gamma$ exceeds the band width for $U/V>\sim 1.6$.
We now comment on the applicability of the BCS mean-field theory to our effective Hamiltonian.
In the context of BCS-BEC crossover, the relation between critical temperature $T_{\rm c}$ and the strength of pairing interaction was extensively studied on the basis of the three-dimensional negative-$U$ Hubbard model.
The calculation of $T_{\rm c}$ for this model at quarter filling was carried out by using the self-consistent $T$-matrix approximation\cite{Keller} and the quantum Monte-Carlo (QMC) simulation.\cite{Sewer}
The result of the QMC study is that $T_{\rm c}$ takes the maximum value $T_{\rm c}^{\ast}\simeq 0.35 t$ at $|U|=|U^{\ast}|\simeq 8t$ (see Fig.~2 in ref.~19).
The BCS result agrees with the QMC result only for $|U|\ll |U^{\ast}|$.
On the one hand, if $|U|$ exceeds the band width, then the BEC formula $T_{\rm c} = t_{\rm B}\times$[numerical constant ($\simeq 3$)], 
where $t_{\rm B}=2t^2/|U|$ is the effective hopping amplitude of composite bosons, becomes to be a good approximation.
Our pairing interaction $\gamma$ is so large that we can not use the BCS mean-field theory to give quantitative predictions.
We here use it to get only qualitative aspects such as relative stability between the singlet and triplet pairings, filling dependence of superconductivity, and so on.
In the estimation of critical temperature, we use the BEC formula instead.

\begin{figure}[t]
\begin{center}
\includegraphics[width=.90\linewidth]{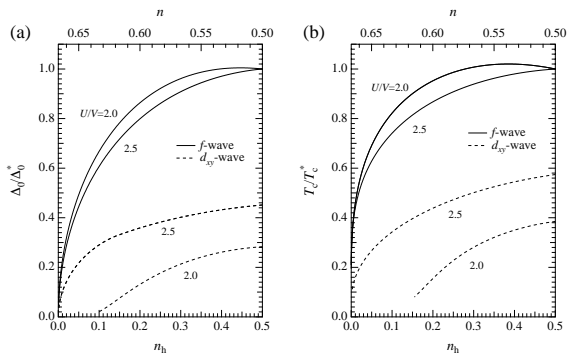}
\end{center}
\caption{Hole-density dependence of (a) zero-temperature order parameter $\Delta_0$ and (b) critical temperature $T_{\rm c}$
for the triplet $f$-wave and singlet $d_{xy}$-wave superconductivities, where the units of order parameter $\Delta_0^{\ast}$
and critical temperature $T_{\rm c}^{\ast}$ are the values for the $f$-wave pairing at quarter-filling, $n_{\rm h}=n=0.5$.
}
\label{f5}
\end{figure}

We turns to the solution of the self-consistent equations (\ref{eq:mfeq1}) and (\ref{eq:mfeq2}).
In Fig.~\ref{f5}, we show the order parameter $\Delta_0=\Delta(T=0)$ and the critical temperature $T_{\rm c}$ for the $f$- and $d_{xy}$-pairings
as a function of hole density of $A$-sublattice, $n_{\rm{h}}\;(=2-3 n)$.
It is found that the introduction of holes leads to rapid development of superconductivity, and $\Delta_0$ and $T_{\rm c}$ saturate at $n_{\rm{h}}\simeq 0.5$.
Thus, the quarter filling, $n=n_{\rm h}=0.5$, is nearly optimal doping rate for superconductivity.
Although the overall behaviors are similar between those two pairings,
$T_{\rm{c}}$ and $\Delta_0$ of $f$-pairing are larger than those of $d_{xy}$-pairing for all $U/V$,
which is mainly coming from the fact that the $d_{xy}$-pairing interaction is smaller than
the $f$-pairing interaction due to the vanishing of two of $\Lambda(\mib{\rho})$ in the $d_{xy}$-wave case.
In order to confirm the stability of the $f$-wave superconductivity, we calculate the temperature dependence of the free energies of the $f$- and $d_{xy}$-wave states per A-sublattice site.
\begin{figure}[t]
\begin{center}
\includegraphics[width=.65\linewidth]{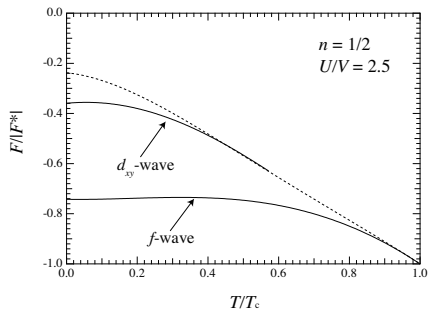}
\end{center}
\caption{Mean-field free energy $F$ for the $f$- and $d_{xy}$-wave pairings as a function of $T$ at quarter filling,
where the temperature is normalized by $T_{\rm c}$ of the $f$-wave pairing and
$F^{\ast}$ represents the value of free energy for the $f$-wave pairing at $T=T_{\rm c}$.
The dotted line represents free energy in case of no pair condensation.}
\label{f7}
\end{figure}
The calculated result for $U/V=2.5$ and quarter filling is shown in Fig.~\ref{f7},
where we find that the free energy of the $f$-wave state is lower than that of the $d_{xy}$-wave state for the whole temperature range.
Thus, we conclude that the $f$-wave superconductivity is stable in the three-fold CO metal.

\begin{figure}[t]
\begin{center}
\includegraphics[width=.64\linewidth]{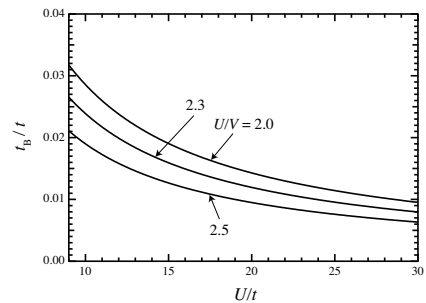}
\end{center}
\caption{Effective hopping amplitude of composite bosons, $t_{\rm B}=2T_1^2/\gamma$,
for the triplet $f$-wave superconductivity as a function of on-site Coulomb interaction $U$.}
\label{f8}
\end{figure}

Finally, we try to estimate $T_{\rm c}$ in the intermediate coupling region, assuming our strong-coupling expansion holds even there.
As mentioned previously, the estimation of $T_{\rm c}$ itself within the BCS theory, which gives unreasonable large values, is meaningless.
Instead, we calculate effective hopping amplitude of composite bosons, $t_{\rm B}=2T_1^2/\gamma$, 
which gives a measure of $T_{\rm c}$ for our effective Hamiltonian.
Using $\gamma$ for the $f$-wave pairing, we show $t_{\rm B}$ as a function of on-site repulsion $U$ in the original Hamiltonian in Fig.~\ref{f8}.
This result suggests $T_{\rm c}\sim 0.01t$ for the superconductivity in the three-fold CO metallic phase.
Also, our strong pairing interaction may cause phase separation near the CO-AF boundary.
This speculation is based on the study of the extended Hubbard ladder by the present authors,~\cite{moro}
where a similar effective Hamiltonian was analyzed by using the Luttinger theory.
They found negative Luttinger parameter ($K_{\rho}<0$), which indicates phase separation, near the CO-AF boundary.

In summary, we have studied the triangular-lattice extended Hubbard model by using strong-coupling expansion.
It has been found that charge carriers in the three-fold CO metal shows the triplet next-nearest-neighbor $f$-wave superconductivity with $T_{\rm c}\sim 0.01t$.
On the one hand, our theory is based on the strong-coupling effective Hamiltonian within the two-hole approximation.
It is desired to study how the omitted many-hole processes affect the present results.
However, the straightforward extension of our perturbation theory is difficult and hopeless.
An alternating way of treating this problem is the VMC method.
Giamarchi {\it et al.} proposed a variational wave function with both the superconducting and magnetic orders in their VMC study of the Hubbard and $t$-$J$ models.~\cite{giamarchi}
In a similar way, we can construct a wave function with both the superconducting and three-fold charge orders. Such investigations are now in progress.

\section*{Acknowledgments}
The authors thank Professor A. Oguchi, Professor H. Yaguchi and Dr. K. Tanaka for valuable discussions.


\begin{thebibliography}{99}
\bibitem{mori}H. Mori, S. Tanaka, and T. Mori: Phys. Rev. B \textbf{57} (1998) 12023.
\bibitem{takada} K. Takada, H. Sakurai, E. Takayama-Muromachi, F. Izumi, R. A. Dilanian, and T. Sasaki: Nature \textbf{422} (2003) 53.
\bibitem{tanaka}Y. Tanaka, Y. Yanase, and M. Ogata: J. Phys. Soc. Jpn. \textbf{73} (2004) 319.  
\bibitem{onari}S. Onari, R. Arita, K. Kuroki, and H. Aoki: Phys. Rev. B \textbf{70} (2004) 094523; \textbf{73} (2006) 014526.
\bibitem{watanabe}H. Watanabe and M. Ogata: J. Phys. Soc. Jpn. \textbf{74} (2005) 2901. 
\bibitem{watanabe2}H. Watanabe and M. Ogata: J. Phys. Soc. Jpn. \textbf{75} (2006) 063702.
\bibitem{kuroki}K. Kuroki: J. Phys. Soc. Jpn. \textbf{75} (2006) 114716. 
\bibitem{hotta}C. Hotta and N. Furukawa: Phys. Rev. B \textbf{74} (2006) 193107.
\bibitem{kaneko} M. Kaneko and M. Ogata: J. Phys. Soc. Jpn. \textbf{75} (2006) 014710.
\bibitem{udagawa} M. Udagawa and Y. Motome: Phys. Rev. Lett. \textbf{98} (2007) 206405. 
\bibitem{fukuyama}M. Inaba, H. Matsukawa, M. Saitoh, and H. Fukuyama: Physica C \textbf{257} (1996) 299. 
\bibitem{Anderson} P. W. Anderson, {\it{The Theory of Superconductivity in the High-$T_c$ Cuprates}} (Princeton University Press, New Jersey, 1997)
\bibitem{ohta}Y. Ohta, K. Tsutsui, W. Koshibae, and S. Maekawa: Phys. Rev. B \textbf{50} (1994) 13594.
\bibitem{vojta} M. Vojta, R. E. Hetzel, and R. M. Noack: Phys. Rev. B \textbf{60} (1999) 8417.
\bibitem{fukumoto} Y. Fukumoto and A. Oguchi: J. Magn. Magn. Mater. \textbf{310} (2007) e96.
\bibitem{onozawa} M. Onozawa, Y. Fukumoto, A. Oguchi, and Y. Mizuno: Phys. Rev. B \textbf{62} (2000) 9648.
\bibitem{hirsch} J. E. Hirsch: Physica C \textbf{158} (1989) 326.
\bibitem{Keller} M. Keller, W. Metzner, and U. Schollw\"{o}ck: Phys. Rev. B \textbf{60} (1999) 3499. 
\bibitem{Sewer} A. Sewer, X. Zotos, and H. Beck: Phys. Rev. B \textbf{66} (2002) 140504. 
\bibitem{moro} S. Morohoshi and Y. Fukumoto: submitted to Phys. Rev. B. 
\bibitem{giamarchi} T. Giamarchi and C. Lhuillier: Phys. Rev. B \textbf{43} (1991) 12943. 
\end{thebibliography}
\end{document}